\documentclass[aps,prb,twocolumn,superscriptaddress,showpacs]{revtex4}
\usepackage{graphicx}
\usepackage{bm}
\usepackage{amsmath}
\begin{document}
\title{Spin Liquid Ground State of the $S=1/2$ Kagome Heisenberg Model}
\author{Simeng Yan}
\affiliation{Department of Physics and Astronomy, University of California, Irvine, CA 92617}
\author{David A. Huse}
\affiliation{Department of Physics, Princeton University, Princeton, NJ 08544}
\affiliation{Institute for Advanced Study, Princeton, NJ 08540}
\author{Steven R. White}
\affiliation{Department of Physics and Astronomy, University of California, Irvine, CA 92617}

\date{\today}

\begin{abstract}
Condensed matter physicists have long sought a realistic two-dimensional (2D)
magnetic system whose ground state is a {\it spin liquid}---a zero temperature
state in which quantum fluctuations have melted away any form of magnetic
order.  The nearest-neighbor $S=\frac{1}{2}$ Heisenberg model on the kagome
lattice has seemed an ideal candidate, but in recent years some approximate
numerical approaches to it have yielded instead a valence bond crystal.  We
have used the density matrix renormalization group to perform very accurate
simulations on numerous cylinders with circumferences up to 12 lattice
spacings, finding instead of the valence bond crystal a singlet-gapped spin liquid with
substantially lower energy that appears to have $Z_2$ topological order.  Our results, through a
combination of very low energy, short correlation lengths and corresponding
small finite size effects, a new rigorous energy bound, and
consistent behavior on many cylinders, provide strong evidence that the 2D
ground state of this model is a gapped spin liquid.

\end{abstract}


\maketitle

\narrowtext

\section{Introduction}

A spin liquid (SL) is a magnetic system which has ``melted'' because of quantum
fluctuations, even at zero temperature.\cite{Balents}  More precisely, it is a spin ground
state with no broken symmetries, namely no preferred spin orientations,
and no frozen valence bond patterns. Spin liquids, besides being exotic and novel
forms of matter, have been suggested as possible routes to high temperature
superconductivity,\cite{pwa} and their unusual topological properties may be useful in
building quantum computers.\cite{ioffe}  A ``Holy Grail'' of condensed matter physics has
been to find a realistic two-dimensional (2D) spin liquid. Spin-liquid-like
behavior is common in one dimension---quantum fluctuations are stronger in
lower dimensions---and some systems with spatial anisotropy, in effect
interpolating between one and two dimensions, are thought to be spin liquids.\cite{q1d}
The strong charge fluctuations associated with being near the Mott transition,
past which there are no localized spins, apparently
can produce a spin liquid even without geometric frustration.\cite{honeycomb}
Strong multispin interactions, such as ring exchanges, are associated with this regime,
and these terms have been included in 2D
Hamiltonians to force spin liquid ground states.\cite{misg98}
Of great interest would be
a spin liquid with a Hamiltonian that is isotropic in both space and spin
directions, with no multispin interactions and preferably only nearest-neighbor exchange
interactions on a simple lattice, where the spin liquid behavior is driven entirely
by frustration.  
Here we present strong evidence for just such a spin liquid.

The Heisenberg $S=\frac{1}{2}$ kagome lattice model (HKLM) has long been thought to be
the ideal candidate. On the experimental side, the material herbertsmithite,
ZnCu$_3$(OH)$_6$Cl$_2$, is a spin-1/2 kagome antiferromagnet showing spin liquid behavior.\cite{e1,e2,e3,e4}
It is thought to be well
modeled by a HKLM Hamiltonian with disorder and an additional Dzyaloshinski-Moriya
interaction. The role of these perturbations in driving the spin liquid behavior is
unclear.\cite{e1,e2,e3,e4}
On the theoretical side, a key problem is that there are no exact or nearly
exact analytical or computational methods to solve infinite 2D quantum lattice
systems. For one dimensional systems, the density matrix renormalization group\cite{White92,White93}
(DMRG), the computational method used in this work, serves in this capacity.
Another ``Holy Grail'' has been to find an accurate and widely applicable
computational method for 2D many-body quantum systems.

The HKLM has been studied with approximate approaches by leading theorists for
decades,\cite{ve} with proposals for spin liquids and for valence bond crystals (VBCs).\cite{Marston}
In the last few years some new numerical evidence has suggested that the HKLM ground state is
{\it not} a spin liquid. Instead, calculations have indicated that it might be a VBC,
with a large 36-site unit cell, called the honeycomb valence bond crystal
(HVBC). First proposed by Marston and Zeng\cite{Marston}, and explored in more detail by
Nikolic and Senthil\cite{Nikolic}, the HVBC was studied with a perturbative series expansion
by Singh and Huse\cite{Singh}, who found the series for the ground state energy to be rapidly converging and to give a
low energy.  Subsequently, Evenbly and Vidal utilized the multiscale
entanglement renormalization ansatz (MERA) on the HKLM to obtain a similar estimate of
the energy of the HVBC\cite{Vidal}. %emerged as
%the ground state.
MERA is related to DMRG but can be implemented directly in
2D.  This MERA study also produced the lowest numerically exact upper bound to the energy
per site, which was close to the series expansion energy.
A recent DMRG study\cite{Sheng}, in
contrast, found a spin liquid ground state, but on the largest lattices the
energy obtained was substantially above that of the HVBC, suggesting that the
method had not found the true ground state.

Here we perform a much more extensive DMRG study of the HKLM with important
differences in techniques from the previous study.  Most importantly, we avoid
fully periodic (toroidal) boundary conditions, which are known to greatly
magnify the truncation errors
in DMRG compared to samples with open ends\cite{White93}.
We present compelling evidence that the ground state of the HKLM is a
spin liquid. This spin liquid has an energy which is substantially below that
of the HVBC.  We also provide a new rigorous upper bound on the ground state energy
of the infinite
2D system. The spin liquid ground state is gapped to both singlet and triplet
bulk excitations, with short correlation lengths.  The short correlation
lengths make the finite circumference (up to 12 bond lengths) cylinders that
we can study converge rapidly towards two dimensions both in energy and
properties.  The spin liquid resembles a $Z_2$ topologically-ordered
short range resonating valence bond (RVB)
state,\cite{rvb,wen,ms,msp} but has significant internal structure with a key role apparently played by eight-site resonant loops.

\section{DMRG methods}

The density matrix renormalization group\cite{White92}  is ideal for one dimensional systems,
but also has long
been used for finite-width 2D strips or cylinders\cite{White98,Cherny}. To maintain a constant accuracy as the width
of the system is increased, the number
of states kept per block $m$ must increase exponentially,
increasing the computer resources needed,
and thus limiting the maximum width feasible.
High accuracy is required to distinguish between phases in the KLHM, so we have
limited the systems studied to a maximum circumference of 12 lattice spacings, for which
we can obtain a relative error of less than 0.1\% in extrapolated energy.

We use a cylindrical geometry for most of our calculations, which we label
by the cylinder's orientation and circumference.  We put the circumference of the cylinder
either along or close in orientation to the $y$-axis.
For example, in our notation YC8 indicates a cylinder (C) in which one of the three bond orientations
is along the $y$-axis (Y), and the circumference is 8 lattice spacings.
For  cylinders oriented with
some bonds along the $x$-direction (X), the circumference is measured in units of $\sqrt{3}/2$ times
the lattice spacing, so that XC8 has a circumference of $c=4 \sqrt{3}$ lattice spacings. For cylinders
which connect periodically around the cylinder's axis with a shift,
the shift is added to the label: e.g. YC9-2 indicates a YC9 cylinder which is connected with
a horizontal shift at the ``seam'' of two units; YC9-2 has a circumference of $c=2\sqrt{21}$.
In these cases with shifts
the kagome lattice is wrapped on the cylinder in a spiral.

There are two key issues that complicate the calculations: 1) the geometry of the cluster---the circumference,
the periodic shift, the boundary conditions at the open ends---can affect the ground and excited states significantly; and 2)
on the wider systems, DMRG can get stuck in metastable states.  Both of these problems can
be surmounted, but it is necessary to study a wide variety of systems and initial states to find
out which geometries do not frustrate the lowest energy states and how to prepare initial
states which permit the algorithm to find the ground state. The results presented here
are a very small fraction of the total number of systems studied, and have been selected to
illustrate the key behavior most clearly.

\section{The valence bond crystal versus the spin liquid}

\begin{figure}[tbp]
\center{
\includegraphics[width=6.5cm]{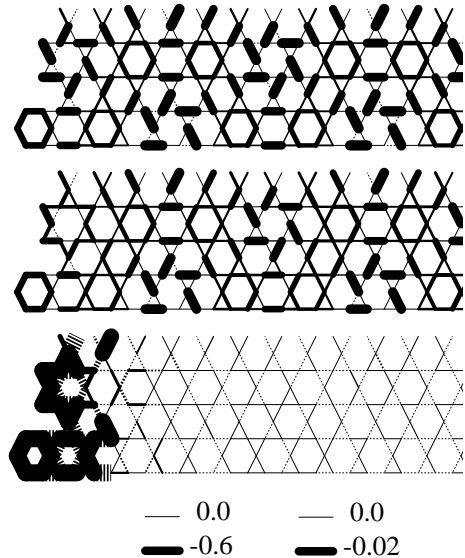}
\caption{Valence bond pattern
for the left two-thirds of a 208 site XC8 cylinder. The width of the lines is proportional to
$|\langle \vec S_i \cdot  \vec S_j \rangle|$ (top two panels and left key) or
$|\langle \vec S_i \cdot  \vec S_j \rangle - e_{\alpha}|$ (bottom panel, right key),
with dashed lines indicating the quantity is positive.
Here $\alpha$ signifies bond direction and the approximate bulk bond
correlations are $e_\alpha  = -0.223$  for horizontal bonds and $-0.217$ for diagonal.
The top panel shows sweep 6 with $m=200$, the middle panel sweep 14 with $m=600$,
and the bottom panel the final sweep 34 with $m=8000$.
Near $m=2400$ (not shown), the energy drops fairly abruptly by about 0.1\%, and the HVBC is
replaced by a spin liquid state.
The line widths have been constrained to a maximum near the left edge, bottom panel.}}
\end{figure}

In this section we present some of the key evidence that the ground state of the KLHM is
a spin liquid, and specifically not the HVBC.
To rule out the HVBC on the cylinders we have studied, despite the possibility of metastable
states, a valuable technique is to favor the HVBC, both in initial state and boundary conditions.
In spite of thus favoring it, the HVBC
is unstable in the resulting
simulations, which is strong evidence against an HVBC ground state.
A typical example is shown in Fig. 1.  In this simulation the cylinder circumference and wrapping vector
accommodate the HVBC state, the left and right edges of the cluster were
trimmed to accommodate and pin the HVBC state, the initial state was prepared in the HVBC state, and
the ordering of the sites of the cylinder used by the DMRG followed an irregular path which always
makes any two sites sharing a valence bond in the HVBC adjacent.  (This biased special ordering allows
a perfect non-resonating HVBC to be represented keeping only $m=2$ states per block.)
These biases towards the HVBC state (particularly the ordering) make it metastable out to
about $m=2400$, after which it transitions to a spin liquid state.
The irregular
edges (along with a variety of other considerations)
allow us to completely rule out the possibility that the uniform state is a superposition of shifted
HVBCs. The final state is a
remarkably uniform spin liquid in the center of the system, with only slight perturbations from the open,
irregular edges and incomplete convergence, and a slight anisotropy from the finite circumference.
This system clearly has a very small length scale for the decay
of the perturbations due to the ends.  
If we use a more conventional ordering of sites, the HVBC is not even metastable;
it disappears within the first few sweeps with $m < 200$.  Energies are generally lower for
a fixed $m$ for the standard ordering than for the special ordering.

We do not find the HVBC to be the ground state on any of the cylinders we have studied.
However, on the largest circumference cylinders, for which our accuracy is much reduced,
the HVBC can be metastable even with the more unbiased standard site ordering.  (A small
bias is still present favoring the less entangled HVBC state.)
For these systems we can compare energies between a prepared
HVBC state and other states. Even in cases where the final energies are not accurate enough
to clearly pick one state, we may be able to judge that the HVBC is not the ground state.
 In Fig. 2 we show such a case for a 400 site YC12 cylinder.
 Two runs were performed, one pinned to start as a HVBC, the other started with an essentially random initial state.
Just after the pinning is released at $m=600$, we see the HVBC has lower energy. However,
near $m=2000$ the initially random state is finding the spin liquid state and becomes lower
in energy.  Subsequently, the difference between the two energies widens as $m$ is increased.
The final state (not shown) appears to be a spin liquid with several localized defects which the simulation
had not yet eliminated.

\begin{figure}[tbp]
\center{
\includegraphics[width=5.5cm]{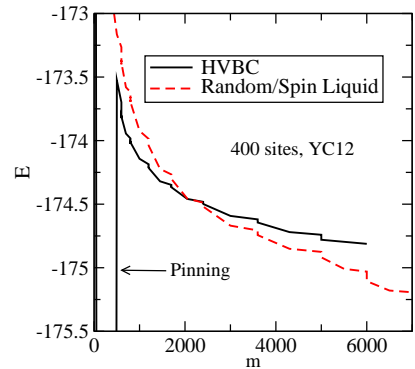}
\caption{Total energy $E$ as the number of states kept $m$ is increased
for a 400 site YC12 cylinder, with two different initial
states---one essentially random, the other with temporary pinning fields to force  an HVBC.
}
}
\end{figure}

The series expansions of Singh and Huse\cite{Singh} treat the HVBC strong-bond interactions as having
a fixed $J=1$, while the weak bonds have their interactions modified by a factor $\lambda$;
the expansion is about $\lambda=0$, and the uniform kagome lattice has $\lambda=1$. We have studied
the ground state along this path of modified Hamiltonians $H(\lambda)$ for a 194-site XC8 cylinder.
Along this path the system has an apparent first-order phase transition near $\lambda_c=0.984$.  This transition
is most clearly seen using a ``hysteresis plot''.  In Fig. 3 we show results from three different
runs.  Each run has a fixed $m$, and $\lambda$ is changed between each DMRG sweep,
with $\lambda$ first increasing, then decreasing 
(note the arrows).
The simulation cannot adapt the wavefunction
from one phase to another in one sweep, so the system shows hysteresis,
staying in the old higher energy state for several sweeps before drifting down to the new lower energy state. The
crossing point of the curves in the two different directions converges very rapidly with increasing $m$ to the
transition point $\lambda_c$, even though the total energies have lower accuracy.  This first-order transition
shows why the series expansion converged so well,\cite{Singh} but to an energy that was not the ground state energy at
$\lambda=1$.

\begin{figure}[tbp]
\center{
\includegraphics[height=6.0cm]{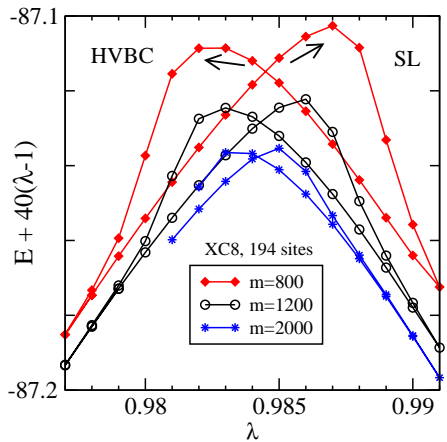}
\caption{Hysteresis plot for a 194 site XC8 cylinder showing the apparent first-order
phase transition between valence bond crystal and spin liquid.
For each of the three runs,  the Hamiltonian parameter $\lambda$ (see text)
is changed for every DMRG sweep, thus tracing out a path through parameter space that resembles
the time evolution of a system under changing $\lambda$.  A linear approximation to
the main trend of the energy $E$ with $\lambda$ has been subtracted out. }
}
\end{figure}

For any of the cylinder geometries we can obtain an estimate of the energy per site for
the infinitely long cylinder by subtracting energies of cylinders of different length to eliminate
end effects. The best results are obtained by doing a sequence of $m$ values for each
cluster and
extrapolating in the truncation error.
For $\lambda=0.98$ on XC8, which is in the HVBC phase,  we obtain  an energy per site of $-0.43237(4)$.
This agrees fairly well with the series expansion energy for this cylinder and $\lambda$, $-0.431(1)$.  This supports
the idea that the series expansion gives a reasonable estimate of the energy of the
HVBC phase at $\lambda=1$ in two dimensions:  $-0.433(1)$,\cite{Singh} as does the MERA HVBC energy,
$-0.4322$,\cite{Vidal} which is a rigorous upper bound.  MERA produces a rigorous upper bound because it
generates a 
wavefunction for the infinite 2D system and evaluates its energy exactly (up to floating point round-off errors).\cite{Vidal}

\section{Ground state energies}

It is possible to generate rigorous
upper bounds on the ground state energy of the infinite 2D system from our results for finite open systems.
Consider an open cluster C which can be ``tiled'' to fill all of 2D, with no sites left out,
and having an even number of sites $N_C$.
We take as a 2D variational
ansatz a product wavefunction, the product being over all the tiles, where we use
our best variational wavefunction for C  (call it $|C\rangle$, with energy $E_C$)
as the ansatz for each tile.
The energy of any of the missing bonds connecting different tiles is zero, since $\langle C|\vec S_i|C\rangle = 0$
for any spin $i$.
Therefore the energy per site of this simple product wavefunction is
$E_C / N_C$.

This approach is crude and converges slowly with the cluster size, with an error proportional to
one over the width.  Nevertheless, the SL energy
 is sufficiently low that we have been able
to obtain a new rigorous upper bound on the 2D energy:  $E_0^{(2D)}<-0.4332$.  This was obtained with
a width-12 open strip (which looks like XC12 unrolled) with $N_C = 576$, keeping $m=5000$ states.   The interior of this
cluster had the uniform valence bond pattern expected for a spin liquid.

\begin{table}[ht]
\caption{
Ground state energies and gaps for infinitely long cylinders of various
circumferences, $c$.
The third column indicates whether the diamond pattern fits perfectly
on the cylinder.\\}
\begin{tabular}{|c|c|c|c|c|c|}
\hline $(c/2)^2$ & Cylinder & DF & E/N & Singlet Gap & Triplet Gap\\
\hline 3 & XC4  & no & $-0.4445$ &  &\\
\hline 4 & YC4   & yes & $-0.4467$ &  &\\
\hline 7 & YC5-2  & no & $-0.43791$ & 0.0108(1)&0.083(1)\\
\hline 9 & YC6  & no & $-0.43914$  & 0.0345(5)& 0.142(1)\\
\hline 12 & XC8 & yes & $-0.43824(2)$ & 0.050(1)& 0.1540(6)\\
\hline 13 & YC7-2 & no  & $-0.43760(2)$  &0.020(1)  &0.055(4)\\
\hline 16 & YC8 & yes & $-0.43836(2)$  &  0.0497(6)&0.156(2)\\
\hline 19 & XC10-1 & no & -0.4378(2) &  &\\
\hline 21 & YC9-2 &  no & $-0.4377(2) $  & 0.032(3)  &0.065(5)\\
\hline 25 & YC10 & no  & $-0.4378(2)$ & 0.041(3)& 0.070(15)\\
%\hline 27 & XC12 &  no  & $-0.4387(16)$  & &\\
\hline 28 & XC12-2  & yes & $-0.4380(3)$  & 0.054(9) &0.125(9) \\
%\hline 31 & YC11-2 & no &   &  & \\
\hline 36 & YC12 & yes & $-0.4379(3)$ & & \\

\hline
\end{tabular}
\end{table}

\begin{figure}[tbp]
\center{
\includegraphics[width=7.0cm]{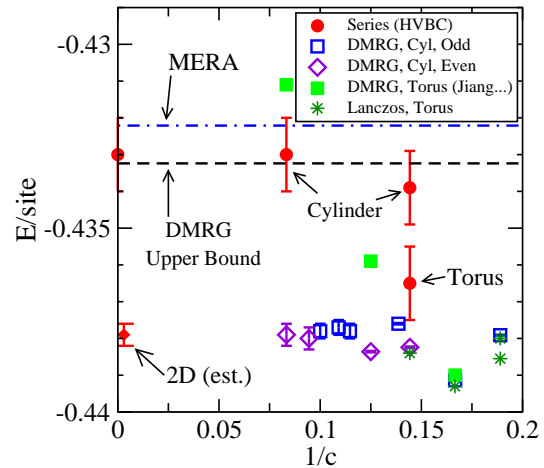}
\caption{Comparison of energies per site for various lattices and methods.
For cylinders, the horizontal axis in this plot is the inverse circumference in
units of inverse lattice spacings. For tori,\cite{Sheng,Leung,aml,ess} the
smallest circumference was used.  In one case we show Lanczos energies for two different
sized (36 and 42 sites) tori that have the same circumference.\cite{aml,ess}
The MERA \cite{Vidal} and our DMRG upper bound results apply directly to
an infinite two
dimensional system, as does the series HVBC result \cite{Singh} that is plotted
on the axis.  The torus DMRG energies\cite{Sheng} are also upper bounds on the true
ground state energies for those tori.
}
}
\end{figure}

Our DMRG results are presented in Table I.  The ground state energies are also plotted and compared to 
other calculations in Fig. 4. 
The DMRG energies are consistent with the Lanczos results \cite{Leung,aml,ess}
and well below the energies of MERA\cite{Vidal} and the series expansions for the HVBC.\cite{Singh}
We note that the previous DMRG result \cite{Sheng} is close to the true ground
state\cite{aml} for the circumference 6 torus.  The entanglement across a cut that separates
a circumference 6 torus into two parts should be roughly the same as across a cut that
separates a circumference 12 cylinder.  We find that circumference 12 is presently our
limit for obtaining good ground state energy estimates on cylinders.  Thus it is perhaps
not surprising that the DMRG results for tori \cite{Sheng} give substantial overestimates
of the ground state energies for circumferences larger than 6.  But these estimates
may alternatively be viewed as variational upper bounds obtained with DMRG.

The XC8 cylinder ($1/c \sim 0.14$) allows a direct comparison
of the energies between the HVBC series and our DMRG:
the DMRG energy is lower by 0.004(1),
and the series result for XC8 is near the 2D result.  The corresponding torus
shows much larger finite size effects in the HVBC series,\cite{Singh} but the true finite size
effects between the tori and cylinders are quite small, as seen by the nearly
identical results from Lanczos on tori and DMRG on cylinders when we use
the largest available torus at each circumference. \cite{Leung,aml,ess}
This is consistent with the small correlation length apparent in Fig. 1.
We conclude that our widest cylinders would have minimal finite size effects
even if the system were in the HVBC phase; in the spin liquid phase, the finite
size effects for the energy are much smaller, and our ground state energies for the
largest cylinders apply to 2D with minimal correction.

We now propose some nonrigorous bounds on the ground state energy of the infinite 2D system, by distinguishing
between frustrated and unfrustrated cylinders.  We call a cylinder unfrustrated if
its ground state energy per site is lower than that of the 2D system.  Here the periodic
boundary conditions do not raise the ground state
energy.  Clearly, most of the very narrow cylinders have low energy
(except possibly YC5-2), presumably due to the contribution to the energy from short RVB resonance
paths that wind around the cylinder.  We also believe that the ``even'' cylinders that are compatible
with the 12-site unit cell diamond pattern that we discuss below are unfrustrated.
Thus we propose that the energy of the XC8 cylinder provides a nonrigorous lower bound
of $-0.4382$ per site on the 2D ground state energy.  This cylinder has the same circumference
as the 36-site torus that is the ``standard'' exact diagonalization cluster (torus).  This ``XC8''
torus has ground state
energy $-0.43838$ per site, \cite{Leung} which is $0.00014(2)$ lower than the XC8
cylinder.  The XC8 cylinder has only 1/3 as many extra winding resonant paths
as does the torus, so should have a correspondingly smaller finite-size effect in its ground state
energy.

We believe that we also have at least one case (and probably many) of a cylinder that is frustrated, with a ground state
energy above that for 2D.  The argument for this also starts with the XC8 cylinder.  Then we cut out
a section of this cylinder with 42 sites and connect it to make a torus.  This torus has various different
new circumferences (topologically nontrivial ways to travel around it), which include
YC7-2 and XC10-1.  We sent this new (42-site!) torus out for exact
diagonalization to Andreas Lauchli who promptly found its ground state
energy, which is approximately 0.0001 per site {\it higher} than that of the XC8 cylinder. \cite{aml}
From this we conclude that one or both of the new circumferences YC7-2 and XC10-1 must
be frustrating to the spin liquid ground state and thus that cylinder should have a higher energy
than the 2D system.  This provides a nonrigorous upper bound on the 2D
ground state energy of $-0.4376$.  Thus our current estimate of the 2D energy
per site is $$E_0^{(2D)}=-0.4379(3)~.$$  Note that all the cylinders wider than YC8 have the same energy
within the uncertainties, and thus rather small finite-size effects on their
ground state energies.

\section{The structure of the spin liquid ground state}

The spin liquid ground state that we find has only short-range correlations
and its ground state energy converges rapidly for finite circumference
cylinders.  We also find, as we discuss below, that there is a nonzero
energy gap for any excitations, including spin-singlet excitations.
The structure of this ground state is apparently some sort of short-range
resonating-valence-bond (RVB) state.\cite{msp}  The smallest resonant loops of singlet
dimers in a nearest-neighbor RVB state on a kagome lattice each surround only one of the
hexagons of the lattice.  The possibilities are tabulated in, e.g., Table I of Nikolic and Senthil.\cite{Nikolic}
If the RVB state had all dimer covers equally weighted, all 32 of these elementary
loops would be equally present.  What we find is that the ground state instead appears to
substantially overweight certain 8-site loops of a diamond (or rhombus) shape.
We first noticed this in seeing what patterns are
easily produced in response to the details of how the samples are cut off at the ends.

\begin{figure}[tbp]
\center{
\includegraphics[width=6.5cm]{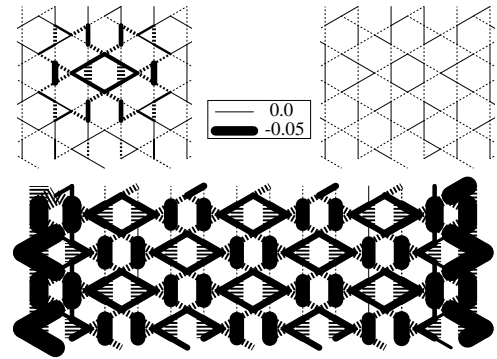}
\caption{Response of the spin liquid to small perturbations on cylinder YC8. In each case certain bonds
have been strengthened: upper left, an 8 site diamond by 1\%; upper right, a 6 site hexagon
by 1\%; and bottom, the wide strong and weak vertical bonds (28 of each) by $\pm 0.5\%$.
Line widths indicate subtracted bond correlations, as in Fig. 1, bottom panel. Surrounding
dimers arise in response to the diamond pinning in the upper left panel; the diamonds arise
in response to dimer pinning in the lower panel.
}
}
\end{figure}

To test the response of the ground state to enhancing each of these elementary resonant loops,
we slightly increased the exchange couplings along the bonds of such a loop at the center
of a YC8 cylinder, and saw how much
this enhanced the spin-spin correlations along the loop and elsewhere.  It is the 8-site diamond
loop that elicits the strongest response, as is shown in Fig. 5.
The 6-site ``perfect hexagon'' loop that has
received attention in previous papers\cite{Marston,Nikolic,Singh} shows a much smaller response, suggesting that
this resonant loop is actually underweighted in the RVB ground state.

One can not tile a kagome lattice with just these favored resonant diamonds.  However,
a particular ``diamond pattern'' valence bond crystal, shown in Fig. 5, appears to be closely related to the spin liquid,
and it is useful to think of the spin liquid as a melted state of this crystal.  One
must keep in mind that the actual overlap of this state with the spin liquid is still small and
the fluctuations are large.  We have measured the correlation length for VBC correlations in this diamond
pattern along cylinder YC8 and find that it is less than 1.5 lattice spacings.  Unlike the HVBC, this
diamond VBC evolves smoothly into the spin liquid without any phase transition as one changes the strengths
of the exchanges appropriately to favor it (we call this ``pinning it'').
For the even cylinders on which this diamond VBC does fit, this
is very useful, as it allows a careful production of the spin liquid, by approaching it in a
controlled and smooth fashion from this diamond VBC.

This diamond pattern was essential in extracting our most accurate ground state energy
estimate for YC12.  Applying the pinning pattern of the lower panel of Fig. 5 reduces
the entanglement of the resulting state, allowing us to obtain more accurate
energies.  The pinning applies an equal number of positive and negative terms, so the energy dependence on the pinning coefficient $\eta$ has no linear term near $\eta=0$
for the uniform spin liquid. After verifying this on several cylinders and determining
that a nearly pure quadratic behavior held for $\eta \le 0.02$, we extrapolated
$E(\eta)$ using the simple formula $E(0) = 4/3 E(0.01) - 1/3 E(0.02)$.
 This procedure reduced our energy
uncertainty for YC12 by a factor of about 5.  (The standard approach of simply
extrapolating to zero truncation error at zero pinning
gave $-0.4375(14)$.)

\begin{figure}[tbp]
\center{
\includegraphics[width=8.5cm]{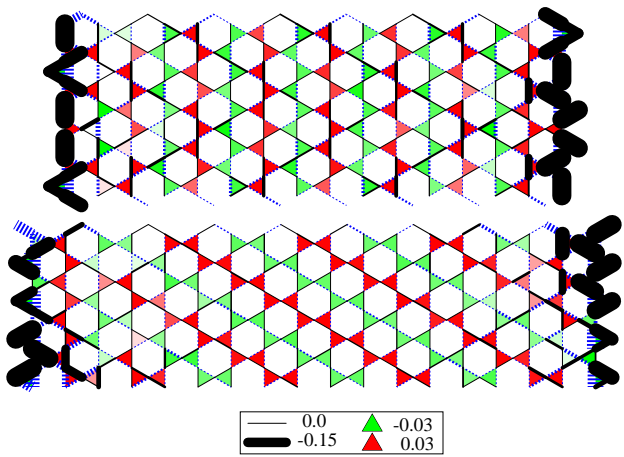}
\caption{Ground state energy patterns for a YC10 (top) and a YC9-2 cylinder. The colors of the
triangles, and their intensities, indicate the deviation of the sum of the
spin-spin correlations on the three bonds forming the triangle
from $3 e_0$, where we take $e_0 = -0.219$. }
}
\end{figure}

The infinitely long cylinders may be viewed as one-dimensional systems with a unit cell
containing multiple spins.  For those ``even'' cylinders that are compatible with the diamond VBC,
this unit cell contains an even number of spins (e.g., for YC8 it contains 12 spins).
In these cases the ground state of the infinite cylinder appears to be nondegenerate
and gapped.  But for the ``odd'' cylinders that are not compatible with the diamond VBC, the
unit cell contains an odd number of spins, and the Lieb-Schultz-Mattis theorem says the
ground state must be degenerate.\cite{lsm}  We find that the ground states on these odd cylinders weakly %dimerized,
break translational invariance, spontaneously
doubling the unit cell, and this produces a pair of degenerate ground states, still with a
gap to higher excited singlet states.   The symmetry breaking is in a ``striped'' pattern
that is shown in Fig. 6.  For YC6 and YC10 the stripes run around the circumference, while for the
other odd spiral cylinders the stripes are spirals.  

\section{Gaps}

To explore the low-lying excited states on our cylinders we use
the following DMRG procedure: first target only one
state, and sweep enough to obtain a high-accuracy ground state.  Then restrict the range
of bonds which are updated in the DMRG sweeps to the central half of the sample,
and now target the two lowest-energy
states, again sweeping to high accuracy, but keeping the end regions of the samples locally in
the ground state.  This restricted sweeping prevents the
low-lying excitations from being bound to the ends of the sample. 
This technique is particularly important for obtaining the singlet gap.  For the
triplet gap, we can also apply magnetic fields on the ends to prevent any edge
excitations from appearing and hiding the bulk gap. For the triplet, we can also
target both states together, one with total $S_z=0$ and the other with $S_z=1$,
or run the two states separately.  These different approaches allowed for fairly independent
checks on the results; in addition, we also varied the lengths and how quickly
the number of states kept was increased.
The results for these gaps are presented in Table I.
Getting gaps is more demanding than getting ground state energies, so our gap estimates
do not go to as wide cylinders as do our ground state estimates.

\begin{figure}[tbp]
\center{
\includegraphics[width=5.5cm]{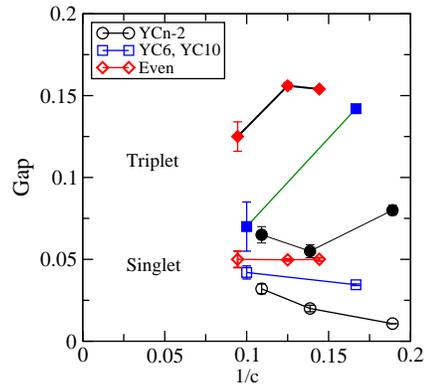}
\caption{Spin triplet (solid symbols) and singlet (hollow symbols) gaps for
various cylinders with circumferences $c$.}
}
\end{figure}

Our gap results for the singlet and triplet gaps differ in an interesting way from
what is known about the gaps from exact diagonalization.\cite{Leung,ED}

We believe that we have found good evidence for a nonzero singlet gap
of about 0.04 or 0.05 in the 2D system.  This is quite different from the exact diagonalization results,
where there are many lower-lying singlets, and the lowest singlet gap is only about 0.01 on the
36-site torus.\cite{ED}  As can be seen in Table I, the singlet gap 
is 0.050, within the errors, for the even XC8 and YC8 cylinders, and it remains near this
value, although with much larger uncertainty, for the wider even XC12-2 cylinder.  The odd cylinders, on the other hand,
have smaller singlet gaps that increase as the circumference is increased, as shown in Fig. 7.  The odd cylinders come in
at least two ``families'': YC6 and YC10 are not spirals, while
YC5-2, YC7-2 and YC9-2 are a series of
spirals with increasing circumference.  In each of these (short) sequences of odd
cylinders the singlet gap increases as the
circumference increases, supporting our conclusion that the singlet gap
remains nonzero in the 2D limit.

The triplet (spin) gap on the standard 36-site torus is 0.164 from exact diagonalization.\cite{Leung}
While XC8 and YC8 have gaps which are quite close to this, our results on other cylinders suggest
that the two dimensional triplet gap is much smaller, as can be seen from Fig. 7.
The triplet excitations appear to be composed of two spinons,
but we cannot resolve whether or not the two spinons bind, although in some cylinders
any binding must be very weak.
This composite nature of the
excitation seems to make the finite size effects  and variation between the cylinders more pronounced.
We do not yet understand the details of these effects.
As in the singlet gap, the spiral odd cylinders 
have the smallest gaps, and the even the largest. 
Note that the triplet gap remains
above the singlet gap in the systems we have studied, and we believe it remains nonzero
in the 2D limit.

\section{Indications of $Z_2$ topological order}

A nearest-neighbor RVB wavefunction is a linear combination of nearest-neighbor singlet
dimers covering the kagome lattice.\cite{msp}  For a kagome lattice wrapped on a cylinder, like
we are studying, such dimer covers are in two topologically distinct classes, and dimer
resonances on finite loops remain in the same topological sector.\cite{ms,msp}  We can force our samples
to be in one or the other of these two topological sectors by choosing how many spins to
leave at each end.  An example of this is apparent for the YC10 cylinder in Fig. 6, where we
force the pattern by the number of spins we leave ``sticking out'' at each end.
For odd cylinders, the two sectors are related by translation along the
length of the cylinder, so are degenerate.  For even cylinders, on the other hand, the
finite circumference $c$ allows the two sectors to have different ground state energies.  The
difference in energy per site should vanish as $\sim\exp{(-c/\xi)}$, where $\xi$ is a
correlation length.  We can measure this splitting for YC8, where it is 0.00069(3)
per site, but for the wider even cylinders we do not yet have an estimate, because
it is more difficult to get a reliable ground state in the
higher-energy topological sector.  It is also worth noting that
on XC8 and YC8 the singlet gap above the higher-energy topological ground state is
substantially smaller than the singlet gap above the overall ground state, which is presumably
a factor in making the higher-energy sector more difficult to work with in DMRG.

A domain wall along the cylinder between regions in the two different topological
ground states is a spinon.  For the odd cylinders, the degeneracy of the two topological
sectors means the spinons are unconfined.  However, two spinons might bind with a finite
binding energy: it appears that this might be happening on cylinder YC10 and thus it might
also occur in the 2D limit.  For the even
cylinders, on the other hand, the spinons are confined
by an effective potential that grows linearly with the distance, because the domain between them
is in the higher-energy topological state.  As a result, the excitation across the spin gap for
the even cylinders must be a bound spinon pair.  It remains to be determined if they
remain bound in the 2D limit.

We suspect that the lowest-lying ``bulk'' singlet excitations that we obtain %find with the method we have used so far
are bound pairs of visons.  It will be interesting to try to test this idea by attempting
to pull apart such a vison pair or to try make a single vison near the end of
a cylinder, but this remains for future research.

\section{Conclusions}

Using very accurate DMRG methods we have found a spin liquid ground state
for the HKLM on cylinders with circumference up to
12 lattice spacings.  The energies are much lower than for the competing valence bond
crystal state.  The combination of low energies and small finite size effects due to
short correlation lengths and nonzero singlet gaps, a new rigorous energy bound,
and a simple picture
for the nature of the spin liquid provide compelling evidence that the infinite 2D system is
a gapped RVB-like spin liquid.
Much remains to be understood concerning this phase, particularly
the detailed structure, exchange statistics and dispersion relations
of the various excitations. It should also be instructive to explore the phase diagram in the vicinity
of this simple nearest-neighbor-only Heisenberg model by changing the
Hamiltonian in various ways in order to find what other phases are nearby
and perhaps to move ``deeper'' into this spin liquid where it might be easier
to study.

We thank Andreas Lauchli for sending us
numerous new exact diagonalization results.
We thank Rajiv Singh for explaining how to convert the published
series results for the 36-site torus and the 2D system to results for cylinders.
We thank Claire Lhuillier for suggesting displaying the energies
of the triangles, as in Fig.6.
We also thank them all again, as well Matthew Fisher, T. Senthil, Erik Sorensen,
Shivaji Sondhi, Guifre Vidal, Xiao-gang Wen, Oleg Tchernyshyov and
Miles Stoudenmire for helpful discussions.
We acknowledge the support of the NSF under grants
DMR-0907500 and DMR-0819860.

\end{document}